\documentstyle[12pt]{article}
\begin{document}
%
%
\def\be{\begin{equation}}
\def\ee{\end{equation}}
\def\bc{\begin{center}}
\def\ec{\end{center}}
\def\bea{\begin{eqnarray}}
\def\eea{\end{eqnarray}}
\def\atu{{\alpha_t^U}}
\def\cm{{\cal M}}
\def\dvc{{\Delta V_{cosm}}}
\def\ov{\overline}
\def\gev{{\rm \; GeV}}
\def\tev{{\rm \; TeV}}
\def\mpl{{M_{\rm P}}}
\def\msu{{M_{\rm SUSY}}}
\def\simlt{\stackrel{<}{{}_\sim}}
\def\tb{{\tan \beta}}
\def\cg{{\cal G}}
\def\str{{\rm \; Str \;}}
\def\mun{{M_{\rm U}}}
\def\zzbar{{(z, \overline{z})}}
\begin{titlepage}
\vspace*{-1cm}
\phantom{bla}
\hfill{CERN-TH/95-292}
\\
\phantom{bla}
\hfill{hep-th/9511086}
\vskip 2.0cm
\begin{center}
{\large \bf Supersymmetry and gauge symmetry breaking
        \\ with naturally vanishing vacuum energy}
\end{center}
\vskip 1.5cm
\begin{center}
{\large Fabio Zwirner}\footnote{On leave from INFN, Sezione di
Padova,
Padua, Italy. Supported in part by the European Union under contract
No.~CHRX-CT92-0004.}
\\
\vskip .5cm
Theory Division, CERN \\
CH-1211 Geneva 23, Switzerland
\end{center}
\vskip 1cm
\begin{abstract}
\noindent
We review the construction of $N=1$ supergravity models
where the Higgs and super-Higgs effects are simultaneously
realized, with naturally vanishing classical vacuum energy
and goldstino components along gauge-non-singlet directions:
this situation is likely to occur in the effective theories
of realistic string models.
\end{abstract}
\vskip 1.5cm
\begin{center}
{\it
Invited talk presented at SUSY--95, Palaiseau, France,
15--19 May 1995}
\end{center}
\vfill{
CERN-TH/95-292
\newline
\noindent
November 1995}
\end{titlepage}
\setcounter{footnote}{0}
\vskip2truecm
\newpage
\section{Motivations}

At the level of dimensionless couplings, the Minimal
Supersymmetric Standard Model (MSSM) is more predictive
than the Standard Model, since its quartic scalar
couplings are related by supersymmetry to the gauge
and the Yukawa couplings (for a review and references
on the theoretical foundations of the MSSM see, e.g.,
\cite{sergio}). The large amount of arbitrariness
in the MSSM phenomenology is strictly related to its
explicit mass parameters, the soft supersymmetry-breaking
masses and the superpotential Higgs mass. Such arbitrariness
cannot be removed within theories with softly broken global
supersymmetry: to make progress, {\em spontaneous}
supersymmetry breaking must be introduced.

To discuss spontaneous supersymmetry breaking in a
realistic and consistent framework, gravitational
interactions cannot be neglected. One is then led to
$N=1$, $d=4$ supergravity, seen as an effective theory
below the Planck scale, within which one can perform
tree-level calculations and study some qualitative
features of the ultraviolet-divergent one-loop quantum
corrections. Of course, infrared renormalization effects
can be studied, but they are plagued by the ambiguities
due to the (ultraviolet) counterterms for the relevant and
marginal operators. To proceed further, one must go to $N=1$,
$d=4$ superstrings, seen as realizations of a fundamental
ultraviolet-finite theory, within which quantum corrections
to the low-energy effective action can be consistently taken
into account, with no ambiguities due to the presence of
arbitrary counterterms.

In recent years, two approaches to the problem have been
followed. On the one hand, four-dimensional string models
with spontaneously broken $N=1$ local supersymmetry have
been constructed \cite{ss}: none of the existing examples is
fully realistic, still they represent a useful laboratory
to perform explicit and unambiguous string calculations.
On the other hand, many studies have been performed
within string effective supergravity theories \cite{gcond}:
the loss in predictivity is compensated by the possibility
of a more general parametrization, including possible
non-perturbative effects that are still hard to handle
at the string theory level. The importance of the problem
and the absence of a fully satisfactory solution are
reflected by the number and the diversity of the
related contributions to this workshop \cite{others,kounnas}.

The generic problems to be solved by a satisfactory
mechanism for spontaneous supersymmetry breaking
can be succinctly summarized as follows.
\begin{itemize}
\item
{\bf Classical vacuum energy.}
The potential of $N=1$ supergravity does not have a definite
sign and scales as  $m_{3/2}^2 M_P^2$, where $m_{3/2}$ is the
(field-dependent) gravitino mass and $M_P \equiv 1 / \sqrt{8
\pi G_N}$ is the Planck mass.
Already at the classical level, one must arrange for the vacuum
energy to be vanishingly small with respect to its natural scale.
\item
{\bf ($m_{3/2}/M_P$) hierarchy.}
In a theory where the only explicit mass scale is the reference
scale $M_P$ (or the string scale $M_S$), one must find a convincing
explanation of why the gravitino mass is at least fifteen orders
of magnitude smaller than $M_P$ (as required by a natural solution
to the hierarchy problem), and not of order $M_P$.
\item
{\bf Stability of the classical vacuum.}
Even assuming that a classical vacuum with the above properties
can be arranged, the leading quantum corrections to the effective
potential of $N=1$ supergravity scale again as $m_{3/2}^2 \, M_P^2$,
too severe a destabilization of the classical vacuum to allow for
a predictive low-energy effective theory.
\item
{\bf Universality of squark/slepton mass terms.}
Such a condition (or alternative but equally stringent ones)
is phenomenologically necessary to adequately suppress
flavour-changing neutral currents, but is not guaranteed
in the presence of general field-dependent kinetic terms.
\end{itemize}
{}From the above list, it should already be clear that the
generic properties of $N=1$ supergravity are not sufficient
for a satisfactory supersymmetry-breaking mechanism. Indeed,
no fully satisfactory mechanism exists, but interesting
possibilities arise within string effective supergravities.
The best results obtained so far have been summarized in the
review talk by Kounnas \cite{kounnas}:
\begin{itemize}
\item
It is possible to formulate supergravity models where the classical
potential is manifestly positive semi-definite, with a continuum of
minima corresponding to broken supersymmetry and vanishing vacuum
energy, and the gravitino mass sliding along a flat direction
\cite{noscale}.
\item
This special class of supergravity models emerges naturally, as
a plausible low-energy approximation, from four-dimensional
string models, irrespectively of the specific dynamical mechanism
that triggers supersymmetry breaking. Due to the special geometrical
properties of string effective supergravities, the coefficient of
the one-loop quadratic divergences in the effective theory, $\str
\cm^2$, can be written as \cite{lhcold,lhc}
\be
\label{genstr}
\str \cm^2 \zzbar = 2 \, Q \, m_{3/2}^2 \zzbar
\, ,
\ee
where $Q$ is a field-independent function, calculable from the
modular
weights of the different fields belonging to the effective low-energy
theory. The non-trivial result is that the only field-dependence of
$\str \cm^2$ occurs via the gravitino mass. Since all
supersymmetry-breaking mass splittings, including those of the
massive string states not included in the effective theory, are
proportional to the gravitino mass, this sets the stage for a natural
cancellation of the ${\cal O} (m_{3/2}^2 \, \mpl^2)$ one-loop
contributions to the vacuum energy. Indeed, there are explicit
string examples that exhibit this feature. If this property can
persist at higher loops (an assumption so far), then the hierarchy
$m_{3/2} \ll \mpl$ can be induced by the logarithmic corrections
due to light-particle loops.
\item
In this special class of supergravity models one naturally obtains,
in the low-energy limit where only renormalizable interactions are
kept, universal mass terms for the MSSM states $(m_0,m_{1/2},\mu,A,
B$ in the standard notation), calculable via simple algebraic
formulae from the modular weights of the corresponding fields
\cite{lhc}.
\end{itemize}
All the above results have been obtained for models where the
goldstino corresponds to a gauge-singlet direction of the
supergravity
gauge group. In the following, we would like to summarize some recent
work \cite{bz,bfz} that extends the above results to models where
the spontaneous breaking of supersymmetry proceeds simultaneously
with the spontaneous breaking of some gauge symmetry.
There are various candidates for the gauge group which could be
broken with supersymmetry (the Standard Model gauge group, some
grand-unified gauge group, some hidden-sector gauge group, \ldots),
but we do not want to be committed here to a specific realization.
In our opinion, such an extension is unavoidable if one wants to
incorporate the full structure of superstring models: singlet
moduli of superstring effective theories are indeed charged under
some gauge group broken near the string scale.

The rest of this contribution is organized as follows. In section~2
we present a toy model that illustrates some general properties of
the mechanism under discussion. In section~3 we discuss two models
with $SU(2) \times U(1)$ breaking. In section~4 we comment on some
possible connections with SUSY GUTs, extended supergravities and
four-dimensional string models.
\section{A toy model}

$N=1$ supergravity models\footnote{Unless otherwise stated, we use
the standard supergravity conventions where $M_P = 1$.} are
characterized by their gauge kinetic function $f$ and K\"ahler
function $\cg$, conventionally decomposed as $\cg = K + \log |w|^2
= - \log Y + \log |w|^2$. Consider the model, based on the K\"ahler
manifold $[SU(1,1)/U(1)]^3$, with
\be
e^{\cg} \equiv {|w|^2 \over Y} = { k^2 \over \left( S + \ov{S}
\right) \left( T + \ov{T} \right) \left( U + \ov{U} \right) }
\, ,
\;\;\;\;\;
(k \ne 0) \, .
\ee
It is clear that, sticking to this field parametrization, we cannot
introduce any linearly realized gauge symmetry. However, by making
the field redefinitions\footnote{Notice the danger of reasoning in
terms of field VEVs and not of physical quantities: in units of
$M_P$, the canonically normalized VEV of $(T + \ov{T})$ is always
equal to 1, even when the VEV of the redefined field $H_1$ is equal
to zero, but the two field representations correspond to the same
physics.}
\be
T = {1 - H_1 \over 1 + H_1} \, ,
\;\;\;\;\;
U = {1 - H_2 \over 1 + H_2} \, ,
\ee
we can write
\be
\label{oldg}
e^{\cg} = { k^2 |1+H_1|^2 |1+H_2|^2 \over 4 \left( S + \ov{S}
\right) \left(1 - |H_1|^2 \right) \left( 1 - |H_2|^2 \right) }
\, .
\ee
The denominator of eq.~(\ref{oldg}) suggests two obvious $U(1)$
symmetries that can be linearly realized on the fields $H_1$ and
$H_2$, but the numerator is not invariant. However, by suitably
modifying the superpotential we can move to a model described by
\be
\label{newg}
e^{\cg} = { k^2 |1+ \sqrt{H_1 \, H_2} |^4 \over \left( S + \ov{S}
\right) \left(1 - |H_1|^2 \right) \left( 1 - |H_2|^2 \right) }
\, ,
\ee
which allows to gauge a $U(1)_X$ with charges $X(S)=0$, $X(H_1)=
-1/2$, $X(H_2)=+1/2$. Choosing for the time being a gauge kinetic
function $f=S$ (this choice is not very important for the
following considerations), one can observe that the K\"ahler
metric is well-behaved in the two regions $|H_1|,|H_2|<1$ or
$|H_1|,|H_2|>1$, that the superpotential is analytic for $H_1
\ne 0$ and $H_2 \ne 0$, and that the Lagrangian is invariant
under the discrete symmetries $(H_1 \to 1/H_1, H_2 \to 1/H_2)$
and $H_1 \leftrightarrow H_2$. It is easy to show that the model
defined above has a positive semi-definite potential ($V_F \ge 0$),
and that the total classical potential $V_0 = V_F + V_D$ is minimized
for arbitrary $|H_1|=|H_2|$ and $S$. To describe the physically
inequivalent vacua, we can use the gauge-invariant VEVs $h \equiv
|H_1| = |H_2|$ and $\theta \equiv \arg (H_1 H_2)$. Considering for
simplicity the vacua with $\theta=0$, and defining $s \equiv (S +
\ov{S})$, the physical mass spectrum can be summarized as follows.
The vector boson and gravitino masses, order parameters for gauge
and supersymmetry breaking, are given by
\be
\label{mvecgratoy}
m_X^2 = {2 h^2 \over s (1-h^2)^2} \, ,
\;\;\;\;\;
m_{3/2}^2 = {k^2 (1+ h)^2 \over s (1-h)^2} \, .
\ee
In the spin-0 sector, there are four physical massless
states, and the only massive one corresponds to ${\rm Re}
\, (H_1 - H_2)$, with mass
\be
\label{mscatoy}
m_0^2 = m_X^2 + m_{3/2}^2 {2 (1 + h^2) (1-h)^2 \over h^2 }
\, .
\ee
In the spin-1/2 sector, there are three physical states, with
masses $m_1^2 = m_{3/2}^2$ and
\be
\label{mfertoy}
m_{2,3}^2 = m_X^2 + m_{3/2}^2  \left[  1 + { (1 + h^2) (1-h)^2
\over 2 h^2 } \right]
\pm {1 + h^2 \over h} m_{3/2} \sqrt{ m_{3/2}^2
{(1 - h)^4 \over 4 h^2 } + m_X^2} \, ,
\ee
and the (canonically normalized) goldstino can be written as
$\tilde{\eta} = (\hat{\tilde{S}} + \hat{\tilde{H}}_1 +
\hat{\tilde{H}}_2) / \sqrt{3}$, where hats denote canonically
normalized fields.

A number of observations are now in order:
\begin{itemize}
\item
Since the goldstino components along $\hat{H}_{1,2}$ are
unsuppressed, the gravitino has interactions of gauge strength
via its $\pm 1/2$ helicity components \cite{fayet}.
\item
$\str \cm^2 = - 10 \, m_{3/2}^2$: this opens the possibility
of cancelling the ${\cal O} (m_{3/2}^2 M_P^2)$ quantum corrections
to the vacuum energy when including other sectors of the full
theory. For example, $n$ scalars with vanishing VEVs and canonical
kinetic terms would give an extra positive contribution $\Delta
\str \cm^2 = 2 \, n \,  m_{3/2}^2$.
\item
The superpotential $w$ has a non-trivial monodromy around $h=0$
($H_1 \to - H_1, H_2 \to - H_2$), a situation already encountered
when studying non-perturbative effects in supersymmetric theories
\cite{npglo}.
\item
In the limit $m_{3/2} \ll m_X$ (which can be reached, e.g. by
choosing $k \ll 1$ and $h$ generic), the effective theory below
the scale $m_X$ would be described by $[H_S \equiv (H_1+H_2)/
\sqrt{2}]$
\be
\label{effg}
e^{\cg} = { k^2 |1+H_S|^4 \over \left( S + \ov{S}
\right) \left(1 - |H_S|^2 \right)^2 }
\, .
\ee
Such an effective theory would not display any singular behaviour
for $h \to 0$, and would give a different value for the coefficient
of the one-loop quadratic divergences, $\str \cm^2 = - 6 \,
m_{3/2}^2$.
This should remind us that a number of problems, such as the
singularity
structure near the cut-off scale and the evaluation of ${\cal O}
(m_{3/2}^2 M_P^2)$ contributions to the vacuum energy, are beyond
the reach of the low-energy effective theory, and need the knowledge
of the full theory to obtain meaningful answers.
\item
In the limit $h \to 0$, one should recover unbroken gauge symmetry
($m_X^2 \to 0$) with broken supersymmetry $(m_{3/2}^2 \to k^2 / s
\ne 0)$, but there are some states whose masses diverge like $1/h$:
\be
\label{split}
m_0^2 \to {2 m_{3/2}^2 \over h^2 } + \ldots \, ,
\;\;\;\;\;
m_2^2 \to { m_{3/2}^2 \over h^2} + \ldots \, .
\ee
This is a signal that, for $h \ll \sqrt{m_{3/2} M_P}$, and denoting
by $\Delta m^2$ the supersymmetry-breaking mass splittings in
eq.~(\ref{split}), the goldstino couplings to the states in
eq.~(\ref{split}) are of order $\Delta m^2 / (m_{3/2} M_P) \sim
m_{3/2} M_P / h^2 \gg 1$: this corresponds to a strongly interacting
goldstino and spoils in general the reliability of perturbation
theory.
\item
Our choice of the gauge kinetic function, $f=S$, was purely
representative, and can be modified while keeping the
result that $\str \cm^2 /  m_{3/2}^2 = {\rm constant}$.
A more general form of $f$ preserving this property is
\be
\label{fgen}
f = \left( S {1 - \sqrt{H_1 H_2} \over 1 + \sqrt{H_1 H_2}}
\right)^{\displaystyle -c/2} \cdot \varphi \left( S {1 + \sqrt{H_1
H_2}
\over 1 - \sqrt{H_1 H_2}} \right) \, ,
\ee
where $c$ is an arbitrary real constant and $\varphi(z)$ is an
arbitrary holomorphic function. The original choice $f=S$ is
recovered for $c=-1$ and $\varphi(z)=\sqrt{z}$. As a curiosity,
observe that, choosing $\varphi(z) = z^{c/2}$, we get
$f=[(1+\sqrt{H_1H_2})/(1-\sqrt{H_1H_2})]^c$. The transformation
$(H_1 \to - H_1, H_2 \to - H_2)$, associated with the monodromy of
$w$ around $h=0$, would correspond in this case to a weak/strong
coupling duality $f \to 1/f$.
\item
Another possibility is to look for different gaugings of the
sigma model under consideration. For example, one could make
the additional field redefinition $S=(1-z)/(1+z)$, and introduce
the superpotential $w = k [ 1 + (zH_1H_2)^{1/3} ]^3$. This would
allow two independent $U(1)$ factors to be gauged, producing a
positive semi-definite potential, broken supersymmetry at all
classical vacua, and less flat directions than in the model defined
by (\ref{newg}). As a candidate form for the gauge kinetic function
$f_{ab}$ ($a,b=1,2$), it is interesting to consider in this case
$f_{ab}= k_a \delta_{ab} \{ [ 1 + (zH_1H_2)^{1/3}] / [ 1 -
(zH_1H_2)^{1/3}]\}^r$, which gives, on the vacua with $z=H_1=H_2
\in {\bf R}^+$, a gaugino mass $m_{1/2} = r \, m_{3/2}$, and has
also interesting properties with respect to weak/strong coupling
duality.
\item
Yet another variant would consist in removing the $S$ field
(either explicitly or by introducing a superpotential that
gives a VEV to its scalar component without giving a VEV to
its auxiliary component), and in assigning to the fields $(H_1
,H_2)$ the K\"ahler potential $K = - (3/2) \log [ ( 1 - |H_1|^2)( 1
- |H_2|^2)]$ and the superpotential $w=k ( 1 + \sqrt{H_1 H_2})^3$.
Choosing $f= L [(1 + \sqrt{H_1 H_2})/(1 - \sqrt{H_1 H_2})]^c$,
with $L$ arbitrary constant and $c \in {\bf R}$, would give a gaugino
mass $m_{1/2}= c \, m_{3/2}$ at all minima with $H_1=H_2 \in {\bf
R}$; the choice $c = \pm 1$ and $L \in {\bf R}$ would guarantee
$m_{1/2}^2 = m_{3/2}^2$ at all minima, corresponding to $|H_1|=
|H_2|$, but would break the discrete invariance under $(H_1 \to
1/H_1, H_2 \to 1 / H_2)$.
\end{itemize}

\section{$SU(2) \times U(1)$ breaking}

Supergravity models of the type considered in the previous section,
with gauge symmetry and $N=1$ supersymmetry both spontaneously
broken, and naturally vanishing classical vacuum energy, can be
systematically constructed by generalizing the previous procedure.
We would like now to discuss two examples in which the broken gauge
group is $SU(2) \times U(1)$, as in the Standard Model.

\begin{center}
{\bf A.}
\end{center}

Consider a model based on the K\"ahler manifold $SU(1,1)/U(1)
\times SU(2,2)/[SU(2) \times SU(2) \times U(1)]$, with the
two factors parametrized by the fields $S$ and by the $2 \times
2$ matrix
\be
Z \equiv \left( \begin{array}{cc}
H_1^0 & H_2^+ \\ H_1^- & H_2^0 \end{array} \right) \, ,
\ee
respectively. We would like to assign, to the degrees of freedom of
$Z$, the $SU(2) \times U(1)$ quantum numbers of the MSSM Higgs
fields,
\be
Z \to e^{i \alpha_A {\tau^A \over 2}} Z
e^{i \alpha_Y {\tau^3 \over 2}} \, .
\ee
In this case, choosing again $f_{ab} = \delta_{ab} S$ for simplicity,
we can introduce the gauge-invariant K\"ahler function
\be
e^{\cg} = {k^2 | 1 + \sqrt{\det Z} |^4 \over \det (1 - Z
Z^{\dagger})}
\, .
\ee
We can easily add to the model a K\"ahler potential and a
superpotential for the squark and slepton sectors, but we shall
omit here this complication. The discussion of the model proceeds
as for the toy model: inequivalent vacua are parametrized by $h$
and $\theta$, there are mass splittings $\Delta m^2 = {\cal O}
(m_{3/2}^2 M_P^2 / h^2)$, and $h$ has to be chosen of the order of
$G_F^{-1/2}$ to correctly reproduce the electroweak scale. This
leads to a dilemma: if the gravitino mass is very light, of order
$h^2/M_P$, then one gets mass splittings of the order of the
electroweak scale, but also an unacceptable tree-level spectrum,
with $\str \cm^2 \simeq 0$ in each mass sector as in global
supersymmetry; if the gravitino mass is of the order of the
electroweak scale, then one gets some huge supersymmetry-breaking
mass splittings, $\Delta m^2 \sim M_P^2$, and the supersymmetric
solution of the hierarchy problem is endangered. Barring possible
string miracles, it would seem that the gauge symmetry breaking
associated with supersymmetry breaking must occur at a much heavier
scale, not too far from $M_P$.

\newpage
\begin{center}
{\bf B.}
\end{center}

An example along this line can be constructed with the K\"ahler
manifold $SU(1,1)/U(1) \times SO(2,n)/[SO(2) \times SO(n)]$,
parametrized by the fields $S$ and $T,H_1,H_2,\ldots$, respectively,
which appears for example in the effective theories of string
orbifold models with tree-level supersymmetry breaking. Consider
the model with $f_{ab} = \delta_{ab} S$,
\be
\label{kah}
K =
- \log ( S + \ov{S} )
- \log [
( T + \ov{T})^2
-
( H_1^0 + \ov{H_2^0} ) ( \ov{H_1^0} + H_2^0 )
-
( H_1^- - \ov{H_2^+} ) ( \ov{H_1^-} - H_2^+ )
-
\ldots]
+ z^{\alpha} \ov{z}_{\alpha} \, ,
\ee
and
\be
\label{www}
w = k
+ {1 \over 2} h_{\alpha \beta}^{(1)} z^{\alpha} z^{\beta} H_1^0
+ {1 \over 2} h_{\alpha \beta}^{(2)} z^{\alpha} z^{\beta} H_2^0
+ \ldots \, ,
\ee
where the superfields $z^{\alpha}$ represent the MSSM quarks and
leptons and we assume, for simplicity, the constants $k$,
$h_{\alpha \beta}^{(1)}$ and $h_{\alpha \beta}^{(2)}$ to be real.

It is easy to show that, for $\langle z \rangle = 0$, $V_F \equiv
0$, and $V_0 = V_F + V_D$ is minimized by $S,T$ arbitrary, $|H_1^0|
=|H_2^0| \equiv h$. With the definitions $s \equiv \langle S+
\overline{S} \rangle$, $t \equiv\langle T+\overline{T} \rangle$
and $x \equiv \langle H_1^0+\overline{H_2^0} \rangle$, the gravitino
mass and the gauge boson masses read
\be
m_{3/2}^2 = \frac{k^2}{s(t^2-|x|^2)}
\ee
and
\be
m_{W,Z}^2 = g_{W,Z}^2 \frac{h^2}{t^2-|x|^2} \, ,
\ee
respectively. Observing that, depending on the relative phase of
$\langle H_1^0 \rangle$ and $\langle H_2^0 \rangle$, $0 \le |x|^2
\le 4 h^2$, we can see that it should be $h^2 / t^2 \simeq m_{W,Z}^2
/ M_P^2$ to reproduce correctly the electroweak scale, irrespectively
of the individual values of $|x|^2$, $h^2$ and $t^2$. In this model,
all the MSSM mass terms depend on the VEVs $s,t,h$ and $x$. To
understand the structure of the model better, we can take the limit
$h / t \rightarrow 0$, which leads to a conventional supergravity
model with hidden sector and, when interactions of gravitational
strength are neglected, to a special version of the MSSM. In such a
limit, the MSSM mass parameters take the special values
\be
\label{masspar}
\begin{array}{ccc}
m_{1/2}^2 = m_{3/2}^2 \, ,
&
m_0^2({\rm matter}) = m_{3/2}^2 \, ,
&
m_0^2({\rm Higgs}) = - m_{3/2}^2 \, ,
\\ & & \\
\mu^2 = m_{3/2}^2 \, ,
&
A^2 = m_{3/2}^2 \, ,
&
B=0 \, .
\end{array}
\ee
Notice the remarkable universality properties, much more stringent
than usually assumed in the general MSSM framework, with one
important exception: since the kinetic terms for the Higgs and matter
fields have different scaling properties with respect to the $t$
modulus, the
corresponding soft scalar masses have different values. In
particular,
the standard mass parameters of the classical MSSM Higgs potential
are given by $m_1^2 = m_2^2 = m_3^2 = 0$, which allows for $SU(2)
\times
U(1)$ breaking already at the classical level, along the flat
direction
$|H_1^0 | = |H_2^0|$.

In summary, we have seen that the structure of the toy model does not
seem suitable for a direct application to $SU(2)_L \times U(1)_Y$
breaking (case A), unless one introduces some extra Standard Model
singlets (case B). Indeed, we know that, in four-dimensional string
models, moduli fields admit points of extended symmetry: in other
words, they are charged under some gauge group broken close to the
string scale. This suggests a second intriguing possibility: to
associate the breaking of supersymmetry with the breaking of a
grand-unified gauge group $G_U$ down to the MSSM gauge group.
Various realizations are possible, depending on the choice of
$G_U$ and of the K\"ahler manifold for the Higgs sector: in this
case we may perform a perturbative study of the dynamical
determination of $M_U$ and $m_{3/2}$, and there may also be
applications to the doublet-triplet splitting problem of SUSY GUTs.
\section{Outlook}

The class of supergravity models discussed in the present talk
has in our opinion rather intriguing properties (including some
formal similarities with recent and less recent results on
non-perturbative phenomena in globally supersymmetric theories),
however it suffers from two main unsatisfactory aspects.
The first is connected with the apparent arbitrariness of the
construction: at the level of $N=1$ supergravity, we are practically
free to choose the gauge group, the number of chiral superfields,
the K\"ahler manifold, the embedding of the gauge group in the
isometry group of the K\"ahler manifold, and finally the gauge
kinetic function and the superpotential that breaks supersymmetry.
The second is connected with the fact that, at the level of $N=1$
supergravity, we are essentially bound to a classical treatment,
given the ambiguities of an effective, non-renormalizable theory
in the control of quantum corrections, both perturbative and
non-perturbative. One may hope to improve in both directions by
establishing some connections with extended $N>1$ supergravity
theories and especially with four-dimensional superstring models.

To obtain a realistic $N=1$ supergravity model, only the candidate
quark and lepton superfields need to transform in chiral
representations of the gauge group. It is then conceivable that
the sector involved in the Higgs and super-Higgs effects can be
obtained, by some suitable projection, from the gauge and
gravitational sectors of an extended supergravity model. Indeed,
spontaneous supersymmetry breaking with vanishing classical
vacuum energy can be associated, in extended supergravities,
with the gauging of a non-compact subgroup of the duality group.
The examples we are aware of give gauge-singlet goldstinos in the
resulting $N=1$ theory, but one could look for models where the
projected $N=1$ goldstino transforms non-trivially under the $N=1$
gauge group: such models would satisfy highly non-trivial
constraints, due to the underlying extended supersymmetry.

Further constraints could be obtained by deriving models of the
type discussed here as low-energy effective theories of
four-dimensional string models with spontaneously broken
$N=1$ supersymmetry. This looks like a natural possibility:
we know many examples of singlet moduli appearing in the
effective string supergravities that are indeed flat directions
breaking an underlying gauge group, restored only at points of
extended symmetry. Unfortunately, the only existing examples
are those in which supersymmetry is broken at the string
tree level, via coordinate-dependent orbifold compactifications:
it should be possible to study these constructions in the cases
where the gauge symmetry and supersymmetry are both spontaneously
broken. This could lead to some progress in the control of
perturbative quantum corrections, since, working at the string
level and not in the effective field theory, we can compute the
full spectrum of states that contribute to the one-loop partition
function.

We hope to return to these problems in some future publication.


\begin{thebibliography}{99}
\bibitem{sergio}
S.~Ferrara, ed., {\em Supersymmetry} (North-Holland,
Amsterdam, 1987).
\bibitem{ss}
C.~Kounnas and M.~Porrati, Nucl. Phys. B310 (1988) 355;
\\
S.~Ferrara, C.~Kounnas, M.~Porrati and F.~Zwirner,
Nucl. Phys. B318 (1989) 75;
\\
M.~Porrati and F.~Zwirner, Nucl. Phys. B326 (1989) 162;
\\
C.~Kounnas and B.~Rostand, Nucl. Phys. B341 (1990) 641;
\\
I.~Antoniadis, Phys. Lett. B246 (1990) 377;
\\
I.~Antoniadis and C.~Kounnas, Phys. Lett. B261 (1991) 369;
\\
I.~Antoniadis, C.~Mu\~noz and M.~Quir\'os, Nucl. Phys. B397 (1993)
515.
\bibitem{gcond}
J.-P.~Derendinger, L.E.~Ib\'a\~nez and H.P.~Nilles, Phys. Lett. B155
(1985) 65;
\\
M.~Dine, R.~Rohm, N.~Seiberg and E.~Witten, Phys. Lett. B156 (1985)
55;
\\
C.~Kounnas and M.~Porrati, Phys. Lett. B191 (1987) 91;
\\
A.~Font, L.E.~Ib\'a\~nez, D.~L\"ust and F.~Quevedo, Phys. Lett.
B245 (1990) 401;
\\
S.~Ferrara, N.~Magnoli, T.R.~Taylor and G.~Veneziano, Phys.
Lett. B245 (1990) 409;
\\
H.-P.~Nilles and M.~Olechowski, Phys. Lett. B248 (1990) 268;
\\
P.~Bin\'etruy and M.K.~Gaillard, Phys. Lett. B253 (1991) 119;
\\
V.~Kaplunovsky and J.~Louis, Nucl. Phys. B422 (1994) 57;
\\
J.H.~Horne and G.~Moore, Nucl. Phys. B432 (1994) 109;
\\
A.~Brignole, L.E.~Ib\`a\~nez and C.~Mu\~noz, Nucl. Phys. B422
(1994) 125 $+$ (E) B436 (1995) 747;
\\
Z.~Lalak, A.~Niemeier and H.-P.~Nilles, Nucl. Phys. B453 (1995)
100 and Phys. Lett. B349 (1995) 99.
\bibitem{others}
J.~Bagger, R.~Brustein, J.-P.~Derendinger, M.K.~Gaillard,
H.-P.~Nilles and T.~Taylor, contributions to these Proceedings,
and references therein.
\bibitem{kounnas}
C.~Kounnas, contribution to these Proceedings, and references
therein.
\bibitem{noscale}
N.-P.~Chang, S.~Ouvry and X.~Wu, Phys. Rev. Lett.
51 (1983) 327;
\\
E.~Cremmer, S.~Ferrara, C.~Kounnas and D.V.~Nanopoulos,
Phys. Lett. B133 (1983) 61;
\\
J.~Ellis, A.B.~Lahanas, D.V.~Nanopoulos and K.~Tamvakis, Phys. Lett.
B134 (1984) 429;
\\
J.~Ellis, C.~Kounnas and D.V.~Nanopoulos,
Nucl. Phys. B241 (1984) 406 and B247 (1984) 373;
\\
S.~Ferrara and A.~Van Proeyen, Phys. Lett. B138 (1984) 77;
\\
U.~Ellwanger, N.~Dragon and M.~Schmidt, Nucl. Phys. B255 (1985) 549;
\\
R.~Barbieri, S.~Ferrara and E.~Cremmer, Phys. Lett. B163 (1985) 143.
\bibitem{lhcold}
S.~Ferrara, C.~Kounnas, M.~Porrati and F.~Zwirner,
Phys. Lett. B194 (1987) 366;
\\
C.~Kounnas, M.~Quir\'os and F.~Zwirner, Nucl. Phys. B302
(1988) 403.
\bibitem{lhc}
S.~Ferrara, C.~Kounnas and F.~Zwirner, Nucl. Phys.
B429 (1994) 589 $+$ (E) B433 (1995) 255.
\bibitem{bz}
A.~Brignole and F.~Zwirner, Phys. Lett. B342 (1995) 117.
\bibitem{bfz}
A.~Brignole, F.~Feruglio and F.~Zwirner, Phys. Lett. B356 (1995) 500.
\bibitem{fayet}
P.~Fayet, Phys. Lett. 70B (1977) 461;
\\
R.~Casalbuoni, S.~De~Curtis, D.~Dominici, F.~Feruglio and R.~Gatto,
Phys. Rev. D39 (1989) 2281.
\bibitem{npglo}
D.~Amati, K.~Konishi, Y.~Meurice, G.C.~Rossi and G.~Veneziano,
Phys. Rep. 162  (1988) 169;
\\
N.~Seiberg and E.~Witten, Nucl. Phys. B426 (1994) 19 and B431 (1994)
484;
\\
N.~Seiberg, Nucl. Phys. B435 (1995) 129.
\end{thebibliography}
\end{document}